\documentclass[twocolumn,showpacs,preprintnumbers,prd,superscriptaddress,nofootinbib]{revtex4-1}

\usepackage{amsmath}
\usepackage{amsfonts}
\usepackage{amssymb}
\usepackage{graphicx}
\usepackage{color}
\usepackage{hyperref}
\usepackage{booktabs}
\usepackage{hyperref}
\usepackage{cleveref}
\usepackage{color}
\Crefrangeformat{equation}{Eqs. (#3#1#4)--(#5#2#6)}
\Crefname{equation}{Eq.}{Eqs.}
\Crefname{figure}{Fig.}{Figs.}
\Crefname{section}{Sec.}{Secs.}

\usepackage{etoolbox}
\makeatletter
\appto{\appendix}{%
  \@ifstar{\def\theequation@prefix{A.}}%
          {}%
}
\makeatother

\begin{document}

\title{Kinematic model-independent reconstruction of Palatini $f(R)$ cosmology}

\author{Salvatore Capozziello}
\email{capozziello@na.infn.it}
\affiliation{Dipartimento di Fisica "E. Pancini", Universit\`a di Napoli ''Federico II'', Via Cinthia, I-80126, Napoli, Italy,}
\affiliation{Istituto Nazionale di Fisica Nucleare (INFN), Sez. di Napoli, Via Cinthia, Napoli, Italy,}
\affiliation{Gran Sasso Science Institute, Via F. Crispi 7,  I-67100, L'Aquila, Italy.}

\author{Rocco D'Agostino}
\email{rocco.dagostino@roma2.infn.it}
\affiliation{Dipartimento di Fisica, Universit\`a degli Studi di Roma ``Tor Vergata'', Via della Ricerca Scientifica 1, I-00133, Roma, Italy.}
\affiliation{Istituto Nazionale di Fisica Nucleare (INFN), Sez. di Roma ``Tor Vergata'', Via della Ricerca Scientifica 1, I-00133, Roma, Italy.}

\author{Orlando Luongo}	
\email{orlando.luongo@lnf.infn.it}
\affiliation{Istituto Nazionale di Fisica Nucleare, Laboratori Nazionali di Frascati, 00044 Frascati, Italy.}
\affiliation{School of Science and Technology, University of Camerino, I-62032, Camerino, Italy.}
\affiliation{Instituto de Ciencias Nucleares, Universidad Nacional Aut́onoma de Ḿexico, AP 70543, Mexico, DF 04510, Mexico}

\begin{abstract}
A kinematic treatment to trace out the form of $f(R)$ cosmology, within the Palatini formalism, is discussed by only postulating the universe homogeneity and isotropy. To figure this out we build model-independent approximations of the luminosity distance through rational expansions. These approximants extend the Taylor convergence radii computed for usual cosmographic series. We thus consider both Pad\'e and the rational Chebyshev polynomials. They can be used to accurately describe the universe late-time expansion history, providing further information on the thermal properties of all effective cosmic fluids entering the energy momentum tensor of Palatini's gravity.
To perform our numerical analysis, we relate the Palatini's Ricci scalar with the Hubble parameter $H$ and thus we write down a single differential equation in terms of the redshift $z$.
Therefore, to bound $f(R)$, we make use of the most recent outcomes over the cosmographic parameters obtained from combined data surveys. In particular our clue is to  select two scenarios, i.e. $(2,2)$ Pad\'e and $(2,1)$ Chebyshev approximations, since they well approximate the luminosity distance at the lowest possible order. We find that best analytical matches to the numerical solutions lead to $f(R)=a+bR^n$ with free parameters given by the set $(a, b, n)=(-1.627, 0.866, 1.074)$ for $(2,2)$ Pad\'e approximation, whereas $f(R)=\alpha+\beta R^m$ with $(\alpha, \beta, m)=(-1.332,  0.749,  1.124)$ for $(2,1)$ rational Chebyshev approximation. Finally, our results are compared with the $\Lambda$CDM predictions and with previous studies in the literature. Slight departures from General Relativity  are also discussed.
\end{abstract}

\maketitle

\section{Introduction}

The cosmological standard model is highly supported by several experimental evidences, among all type Ia Supernovae (SN~Ia) \cite{Perlmutter98,Riess98,Schmidt98}, Baryon Acoustic Oscillations (BAO) \cite{Eisenstein98} and the analysis of Cosmic Microwave Background (CMB) anisotropies \cite{WMAP9,Planck15}. The standard model, in particular, shows up a preferable spatially flat universe undergoing an anomalous speed up at current time \cite{Haridasu17}. In the framework of Einstein's theory of General Relativity (GR), the late-time acceleration would originate from an unknown component dubbed dark energy \cite{Bamba12,Joyce16,Kleidis16}. Dark energy is supposed to provide a negative pressure, with constant equation of state. The simplest candidate for dark energy is the cosmological constant $\Lambda$ \cite{Sahni00,Copeland06} originating from early-time quantum fluctuations of vacuum. The cosmological constant, along with the cold dark matter component, defines the standard paradigm named $\Lambda$CDM model. The problems related to the physical nature of $\Lambda$ have evaded our understanding so far\footnote{Although very successful in accounting for all the major cosmological observables, the $\Lambda$CDM model is plagued by fundamental issues, such as the \emph {coincidence problem} \cite{Zlatev99} and the \emph{fine-tuning problem} \cite{Weinberg89}.}. This fact pushes cosmologists to explore alternative interpretations for the accelerated scenario. For instance, plausible landscapes include dynamical dark energy with evolving scalar fields \cite{Peebles88,Padmanabhan02,Singh03} or evolving equation of state \cite{Chevallier01,Linder03,Jassal05}, attempts to unify dark matter and dark energy into a single fluid \cite{Bento02,Capozziello18}, and higher-dimensions braneworld models \cite{Csaki00,Maartens04}.

An alternative view is to modify GR on cosmological scales. This turns out to explain the late-time acceleration without the need of dark energy. One of the most studied extensions of GR is represented by $f(R)$ gravity \cite{Sotiriou10,Capozziello11,Nojiri17}, generalizing the Einstein-Hilbert action with higher-order curvature terms in the Lagrangian.
Many theoretical studies carried out so far have focused on the cosmological viability of such theories \cite{Nojiri03,Carroll04,Olmo05,Amendola07}.
Also, from the observational point of view, the viability of these models has been tested by means of several cosmological data surveys \cite{Capozziello03,Koivisto06,Santos08,Basilakos13,Nunes17,new1,new2}.

Standard approaches toward the study of $f(R)$ paradigms postulate the $f(R)$ functions \cite{Capozziello06,Fay07,Tsujikawa08,Pires10}. This approach is plagued by the fact that $f(R)$ is assumed \emph{a priori} without relying on first principles. For these reasons, we here implement the inverse procedure, i.e. we start from data and we back-reconstruct the $f(R)$ action without any \emph{ad hoc} assumptions. Reconstructions of $f(R)$ from the dynamics of the universe have been discussed in the context of the metric formalism \cite{Capozziello05,Capozziello17}. In this work, we extend the method to the Palatini formalism by means of the \emph{cosmographic technique}.
In particular, our approach is built upon rational polynomial approximations, which are able to extend the radius of convergence of the cosmographic series and minimize the relative uncertainties in estimating kinematic parameters \cite{Gruber14,Aviles14,D'Agostino18,Anjan}.
We apply this method to find a model-independent expression of the Hubble expansion rate that can be then used to obtain the redshift as a function of the Ricci curvature. Thus, we infer the $f(R)$ action through a numerical integration with initial conditions set by observational constraints on the solar system.

The paper is organized as follows. \Cref{sec:theory} is dedicated to a review of the $f(R)$ gravity models in the Palatini formalism. In \Cref{sec:cosmography} we present the method of rational approximations in the context of cosmography. In \Cref{sec:results} we reconstruct the $f(R)$ action and compare our results with the predictions of the $\Lambda$CDM model and with previous studies in the literature. Finally, in \Cref{sec:conclusion} we summarize results and conclude.

\section{Palatini $f(R)$ cosmology}
\label{sec:theory}

The action describing the $f(R)$ gravity models can be written as \cite{Capozziello02,Allemandi04,Carloni05}
\begin{equation}
S=\dfrac{1}{2\kappa}\int d^4x\ \sqrt{-g}\ f(R) + S_m\ ,
\label{eq:action}
\end{equation}
where $\kappa\equiv8\pi G$ and $G$ is the Newton's constant; $g$ is the metric determinant and $S_m$ is the matter action.
Differently from standard GR, for a non-linear Lagrangian density in $R$ the field equations that one obtains from the least action principle depend on the variational principle adopted. In the Palatini formalism, the action is varied with respect to both metric $g_{\mu\nu}$ and affine connections $\Gamma_{\mu\nu}^\alpha$, which are treated as independent variables. Varying \Cref{eq:action} with respect to $g_{\mu\nu}$ gives:
\begin{equation}
F(R) R_{\mu\nu}-\dfrac{1}{2}f(R) g_{\mu\nu}=\kappa T_{\mu\nu}\ ,
\label{eq:var wrt metric}
\end{equation}
where $F\equiv df/dR$ and $T_{\mu\nu}$ is the energy-momentum tensor defined as:
\begin{equation}
T_{\mu\nu}=-\dfrac{2}{\sqrt{-g}}\dfrac{\delta S_m}{\delta g^{\mu\nu}}\ .
\end{equation}
On the other hand, varying \Cref{eq:action} with respect to $\Gamma^{\alpha}_{\mu\nu}$ provides \cite{Vollick03}:
\begin{equation}
\nabla_\lambda(F(R)\sqrt{-g}g^{\mu\nu})=0\ ,
\label{eq:var wrt connection}
\end{equation}
where $\nabla_\lambda$ is the covariant derivative with respect to the connections. From \Cref{eq:var wrt connection},  one can define the conformal metric $h_{\mu\nu}\equiv F g_{\mu\nu}$, so that the connections become the Christoffel symbols of the metric $h_{\mu\nu}$.  One thus obtains:
\begin{equation}
\Gamma_{\mu\nu}^{\alpha}=\tilde{\Gamma}^\alpha_{\mu\nu}+\dfrac{1}{2F}\left[2\delta^\alpha_{(\mu}\partial_{\nu)}F-g_{\mu\nu}g^{\alpha\sigma}\partial_\sigma F\right] ,
\end{equation}
where $\tilde{\Gamma}^\alpha_{\mu\nu}$ are the Christoffel symbols of the metric $g_{\mu\nu}$. Further, the Ricci tensor of the conformal metric can be written as the sum of the Ricci tensor of the metric $g_{\mu\nu}$, $\tilde{R}_{\mu\nu}$, plus additional terms:
\begin{equation}
R_{\mu\nu}=\tilde{R}_{\mu\nu}+\dfrac{3}{2}\dfrac{(\nabla_\mu F)(\nabla_\nu F)}{F^2}-\dfrac{\nabla_\mu\nabla_\nu F}{F}-\dfrac{g_{\mu\nu}}{2}\dfrac{\Box F}{F}\ ,
\end{equation}
where $\square\equiv \nabla_\alpha\nabla^\alpha$. The Ricci scalar of the conformal metric becomes:
\begin{equation}
R=\tilde{R}+\dfrac{3}{2}\dfrac{(\nabla_\mu F)(\nabla^\mu F)}{F^2}-3\dfrac{\Box F}{F}\ ,
\label{eq:Ricci scalar}
\end{equation}
where $\tilde{R}=g^{\mu\nu}\tilde{R}_{\mu\nu}$.

To obtain the cosmological solutions, we consider the flat Friedmann-Lema{\^i}tre-Robertson-Walker (FLRW) metric\footnote{We work in units of $c=1$.}:
\begin{equation}
ds^2=-dt^2+a(t)^2\delta_{ij}dx^idx^j\ ,
\end{equation}
where $a(t)$ is the cosmic scale factor. The energy-momentum tensor for a perfect fluid of density $\rho$ and pressure $p$ is given by
\begin{equation}
T_{\mu\nu}=\text{diag}(-\rho,p,p,p)\ ,
\end{equation}
and, thus, the trace of \Cref{eq:var wrt metric} results in
\begin{equation}
R F(R)-2 f(R)=\kappa(3p-\rho)\ .
\label{eq:trace}
\end{equation}
From the combination of the $(0,0)$ and $(i,i)$ components of the field equations, one obtains the generalized Friedmann equation:
\begin{equation}
\left(H+\dfrac{1}{2}\dfrac{\dot{F}}{F}\right)^2=\dfrac{1}{6}\left[\dfrac{\kappa(\rho+3p)}{F}+\dfrac{f}{F}\right].
\label{eq:gen Friedmann}
\end{equation}
Taking the time derivative of \Cref{eq:trace} and combining it with the energy conservation leads to
\begin{equation}
\dot{R}=-\dfrac{3H(RF-2f)}{F_R-F}\ ,
\label{eq:dot R}
\end{equation}
where $F_R\equiv dF/dR=d^2f/dR^2$.
Assuming that the universe is filled with pressureless matter and neglecting the contribution of radiation, we have $p=0$ and $\rho=\rho_m$. Since $\dot{F}=F_R\dot{R}$, one can combine \Cref{eq:dot R} with \Cref{eq:gen Friedmann} to finally get:
\begin{equation}
H^2=\dfrac{1}{6F}\dfrac{2\kappa\rho_m+RF-f}{\left[1-\dfrac{3}{2}\dfrac{F_R(RF-2f)}{F(RF_R-F)}\right]^2}\ ,
\label{eq:Friedmann}
\end{equation}
which represents the first modified Friedmann equation in the Palatini formalism.

The physical solution in Palatini's gravity is simple to interpret. The gravitational part of the action may be mapped into a formally equivalent Brans-Dicke theory, as a consequence of replacing metric with connection in the action. This is due to the fact that the
independent connection can be interpreted as an auxiliary field. Hence, after cumbersome algebra, one immediately gets that Palatini's gravity corresponds to a Brans-Dicke model with potential in which one fixes the free parameter to $\omega_0=-\frac{3}{2}$. The most general case of $f(R)$ gravity is not indeed the Palatini formalism, but the metric-affine version of $f(R)$ gravity. Imposing further assumptions can lead to both metric or Palatini scenarios, respectively for $\omega_0 = 0$ and $\omega = -\frac{3}{2}$. In such a scheme one can also notice that Palatini $f(R)$ gravity is a metric theory following the classification made by \cite{Will81}. This reinforces the idea that the independent connections becomes an auxiliary field. In such a way it is not possible to completely frame a metric-affine $f(R)$ gravity with the Palatini formalism. For our purposes, we want to investigate the numerical reconstructions of Palatini $f(R)$ with cosmography. To do so, since Eq. \eqref{eq:trace} with vanishing source on the right provides a solution $\propto R^2$, we may look for polynomial versions of the test functions that we employ in the next sections to extract the form of $f(R)$.

%%%%%%%%%%%%%%%%%%%%%%%%%%%%%%%%%%%%%%%%%%%%%%%%%%%%%%%%%%

\section{Cosmography with rational approximations}
\label{sec:cosmography}

The study of the universe's dynamics can be done through a purely kinematic approach by means of the cosmographic technique \cite{Saini00,Visser,Cattoen07,Carvalho11,Luongo11}. With the only assumption of homogeneity and isotropy on large scales as required by the cosmological principle, cosmography allows one to frame the universe's expansion history at late times with no need of postulating any \emph{a priori} cosmological model.
This method is built upon the Taylor expansion of the scale factor around present time $t_0$ \cite{Aviles12,Busti15}:
\begin{equation}
a(t)=1+\sum_{k=1}^{\infty}\dfrac{1}{k!}\dfrac{d^k a}{dt^k}\bigg | _{t=t_0}(t-t_0)^k\ .
\label{eq:scale factor}
\end{equation}
From this, one can define the cosmographic series:
\begin{align}
&H(t)\equiv \dfrac{1}{a}\dfrac{da}{dt} \ , \hspace{1cm} q(t)\equiv -\dfrac{1}{aH^2}\dfrac{d^2a}{dt^2}\ ,  \label{eq:H&q} \\
&j(t) \equiv \dfrac{1}{aH^3}\dfrac{d^3a}{dt^3} \ , \hspace{0.5cm}  s(t)\equiv\dfrac{1}{aH^4}\dfrac{d^4a}{dt^4}\ .   \label{eq:j&s}
\end{align}
Then, the luminosity distance as function of the redshift reads:
\begin{align}
d_L(z)&=\dfrac{1}{H_0}\bigg[z+\dfrac{1}{2}(1-q_0)z^2-\dfrac{1}{6}(1-q_0-3q_0^2+j_0)z^3 \nonumber \\
&+\dfrac{1}{24}(2-2q_0-15q_0^2-15q_0^3+5j_0+10q_0j_0+s_0)z^4 \nonumber \\
&+\mathcal{O}(z^5)\bigg] \ .
\label{eq:dL}
\end{align}
The Hubble rate in terms of the cosmographic parameters is obtained as:
\begin{equation}
H(z)=\left[\dfrac{d}{dz}\left(\dfrac{d_L(z)}{1+z}\right)\right]^{-1}\ .
\label{eq:H(z)}
\end{equation}

The method of Taylor approximations is unfortunately limited by the short convergence radius, $z<1$. A possible way to perform cosmological analyses at higher redshift domains is to consider rational approximations. Two relevant examples in this respect are the Pad\'e polynomials and ratios of Chebyshev polynomials, which offer clear convergence improvements and significant reductions of error propagation \cite{Gruber14,Aviles14,D'Agostino18}.

\subsection{The  Pad\'e approximations}

In the case of Pad\'e approximations, one starts from the Taylor expansion of a generic function, $f(z)=\sum_{i=0}^\infty c_i z^i$, with $c_i=f^{(i)}(0)/i!$.

One thus defines the generic $(n,m)$ Pad\'e approximant of $f(z)$ by \cite{Litvinov93,Baker96}:
\begin{equation}
P_{n,m}(z)=\dfrac{\displaystyle{\sum_{i=0}^{n}a_i z^i}}{\displaystyle{\sum_{j=0}^{m}b_j z^j}}\,.
\label{eq:def Pade}
\end{equation}
The coefficients $a_i$ and $b_j$ are determined by requiring $f(z)-P_{n,m}(z)=\mathcal{O}(z^{n+m+1})$:
\begin{equation}
\left\{
\begin{aligned}
&a_i=\sum_{k=0}^i b_{i-k}\ c_{k} \ ,  \\
&\sum_{j=1}^m b_j\ c_{n+k+j}=-b_0\ c_{n+k}\ , \hspace{0.5cm} k=1,\hdots, m \ .
\end{aligned}
\right .
\end{equation}

\subsection{The rational Chebyshev approximations}

The second method deals with first kind Chebyshev polynomials defined by \cite{Gerald03}:
\begin{equation}
T_n(z)=\cos(n\theta)\ ,
\end{equation}
where $n\in\mathbb{N}$ and $\theta=\arccos(z)$.
They are orthogonal polynomials with respect to the weighting function $w(z)=(1-z^2)^{-1/2}$ in the domain $z\in[-1,1]$ and bid to the recurrence relation
\begin{equation}
T_{n+1}(z)=2zT_n(z)-T_{n-1}(z)\ .
\end{equation}
Thus, one can write the Chebyshev series of a generic function $f(z)$ as
\begin{equation}
f(z)=\sum_{k=0}^{\infty}{}^{\prime} c_k T_k(z)\ ,
\label{Cheb series}
\end{equation}
where $\sum^{\prime}$ means that the first term in the sum must be divided by 2. The coefficients $c_k$ are thus obtained as:
\begin{equation}
c_k= \dfrac{2}{\pi}\displaystyle\int_{-1}^{1}g(z)\ T(z)\ w(z)\ dz\ ,
\end{equation}
where $g(z)$ is the Taylor series of $f(z)$ around $z=0$. Therefore, applying a similar procedure as well as in the Pad\'e framework, we build the $(n,m)$ rational Chebyshev approximant of $f(z)$ by:
\begin{equation}
R_{n,m}(z)=\dfrac{\displaystyle{\sum_{i=0}^n}{}^{\prime}\ a_i T_i(z)}{\displaystyle{\sum_{j=0}^m}{}^{\prime}\ b_j T_j(z)}\ .
\label{eq:def ratio Cheb}
\end{equation}
In this case, the coefficients $a_i$ and $b_j$ are obtained through:
\begin{equation}
\left\{
\begin{aligned}
&a_i=\dfrac{1}{2}\sum_{j=0}^{m}{}^{\prime}\ b_j(c_{i+j}+c_{|i-j|})=0 \ , \hspace{0.5cm} i=0,\hdots,n   \\
&\sum_{j=0}^{m}{}^{\prime}\ b_j(c_{i+j}+c_{|i-j|})=0 \ , \hspace{0.5cm} i=n+1,\hdots,n+m\ .
\end{aligned}
\right .
\end{equation}

It is possible to generalize the above formalism to consider an arbitrary interval $[a,b]$ of $z$. This can be done by defining the generalized Chebyshev polynomials $T^{[a,b]}_n(z)=\cos(n\theta)$
in terms of the new variable
\begin{equation}
z=\dfrac{a(1-\cos\theta)+b(1+\cos\theta)}{2}\ ,
\end{equation}
obtained through the transform
\begin{equation}
\cos\theta=\dfrac{2z-(a+b)}{b-a}\ .
\end{equation}
In so doing, $\theta\in[-\pi,\pi]$ while $z\in[a,b]$.  Therefore, the generalized Chebyshev polynomials are found from the ordinary Chebyshev polynomials as
\begin{equation}
T_n^{[a,b]}(z)=T_n\left(\dfrac{2z-(a+b)}{b-a}\right) .
\end{equation}
Polynomials $T_n^{[a,b]}(z)$ are orthogonal with respect to the weighting function \cite{Obsieger13}
\begin{equation}
w_{[a,b]}(z)=[(z-a)(b-z)]^{-1/2}\ ,
\end{equation}
so that the inner product is defined as
\begin{equation}
\langle T_m^{[a,b]},T_n^{[a,b]}\rangle=\int_a^b dz\ w_{[a,b]}(z)\ T_n^{[a,b]}(z)\ T_m^{[a,b]}(z)\ .
\end{equation}
Since $T_n^{[a,b]}(z)=\cos(n\theta)$ and $d\theta=-w_{[a,b]}dz$, the orthogonality condition becomes
\begin{equation}
\langle T_m^{[a,b]},T_n^{[a,b]}\rangle=
\begin{cases}
\pi\ , & n=m=0 \vspace{0.2cm}\\
\dfrac{\pi}{2} \delta_{nm}\ , & \text{otherwise}
\end{cases}
\end{equation}

\section{Reconstructing the $f(R)$ action through cosmography}
\label{sec:results}

In this section, we present the method to reconstruct the $f(R)$ function from the Hubble expansion rate $H(z)$ in the FLRW universe.
To do so, we write the relation between the Ricci scalar and the Hubble parameter in the metric formalism:
\begin{equation}
\tilde{R}=6(\dot{H}+2H^2)\ .
\label{eq:R-H metric}
\end{equation}
For practical purposes, we convert the time derivative into derivative with respect to the redshift according to $d/dt=-(1+z)H(z)d/dz$.
Hence, \Cref{eq:R-H metric} can be written as
\begin{equation}
\tilde{R}(z)=-6(1+z)H(z)H'(z) +12H(z)^2\ ,
\end{equation}
where the symbol `prime' denotes derivative with respect to $z$.  Plugging the above expression into \Cref{eq:Ricci scalar}, one finds \cite{Baghram09}:
\begin{align}
R(z)&=12H^2-6(1+z)H\left(\dfrac{HF'+FH'}{F}\right) \nonumber  \\
&+3(1+z)^2\left[\dfrac{2HF(HF''+H'F')-H^2{F'}^2}{2F^2}\right]
\label{eq:R-H Palatini}
\end{align}
Then, combining \Cref{eq:Friedmann,eq:R-H Palatini} and expressing the matter density in terms of $z$ as $\kappa\rho_m=3H_0^2\Omega_{m0}(1+z)^3$, we obtain a differential equation for $F(z)$:
\begin{align}
&F''-\dfrac{3}{2}\dfrac{{F'}^2}{F}+\left(\dfrac{H'}{H}+\dfrac{2}{1+z}\right)F'-\dfrac{2H'}{H(1+z)}F \nonumber \\
&+3\Omega_{m0}(1+z)\left(\dfrac{H_0}{H}\right)^2=0\ .
\label{eq:F-z}
\end{align}
Therefore, once $H(z)$ is extracted from data, we can invert \Cref{eq:R-H Palatini} to find $z$ in terms of $R$ and substitute in the solution of \Cref{eq:F-z} to find $F(R)$. Thus, integrating numerically $F(R)$, we can definitively infer the $f(R)$ function, without any further assumptions made \emph{ad hoc} on its form.

\subsection{Cosmographic outcomes}

In view of the treatment proposed in \cite{D'Agostino18}, we apply the above method to the $(2,2)$ Pad\'e and the $(2,1)$ rational Chebyshev approximations. We report  in the Appendix the expressions of $d_L(z)$, from which one can derive the corresponding $H(z)$ by means of \Cref{eq:H(z)}. In what follows, we fix $\Omega_{m0}=0.3$\footnote{This choice is perfectly consistent with the results of the \textit{Planck} collaboration \cite{Planck15}, within the $1\sigma$ errors: $\Omega_m=0.308 \pm 0.012$\ .}.
As far as the cosmographic parameters are concerned, we assume the best-fit results obtained in \cite{D'Agostino18} through a comparison with the most recent cosmological data surveys.
Those results are quite different from the outcomes that one could get from the standard cosmological model. However, fixing $q_0, j_0$ and $s_0$ to the values that one obtains from the $\Lambda$CDM model would imply the dependence on the standard paradigm. On the contrary, in our analysis we adopt the results we obtained in \cite{D'Agostino18} through a model-independent procedure by using both Chebyshev and Pad\'e treatments.

\begin{figure}
\begin{center}
\includegraphics[width=3.2in]{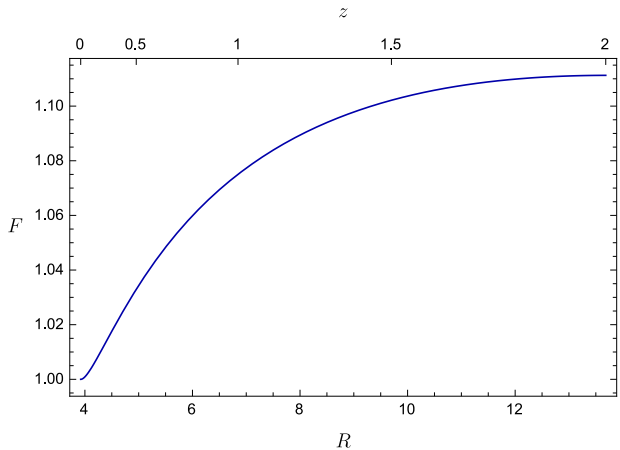}
\caption{Pad\'e reconstruction of $F(R)$ in the redshift interval $[0,2]$.}
\label{fig:F(R) Pade}
\end{center}
\end{figure}

\begin{figure}
\begin{center}
\includegraphics[width=3.2in]{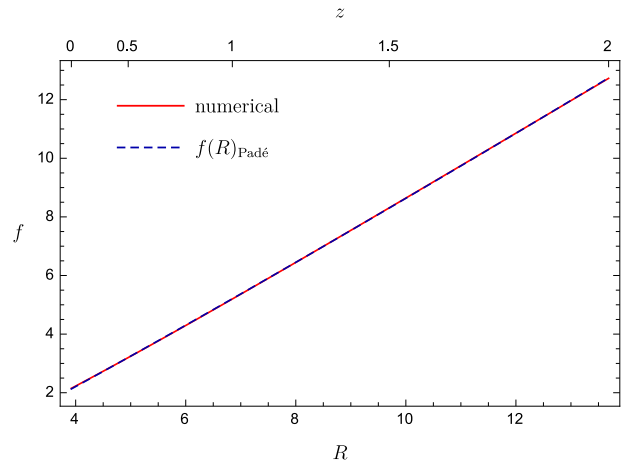}
\caption{Comparison between the Pad\'e reconstruction of $f(R)$ and its best analytical approximation (cf. \Cref{eq:best f(R) Pade}).}
\label{fig:f(R) Pade}
\end{center}
\end{figure}

In the case of the Pad\'e approximant, we consider:
\begin{equation}
\left\{
\begin{aligned}
&h_0= 0.6494^{+0.0211}_{-0.0202}\ ,\\
&q_0=-0.285^{+0.040}_{-0.046}\ , \\
&j_0=0.545^{+0.087}_{-0.084}\ , \\
&s_0=0.118^{+1-135}_{-1.025}\ ,
\end{aligned}
\right.
\label{cosm param Pade}
\end{equation}
where $h_0\equiv H_0/(100\ \text{km/s/Mpc})$ is the dimensionless Hubble constant.
The initial conditions to solve \Cref{eq:F-z} are found by requiring that the effective gravitational constant, $G_\text{eff}=G/F$, is equivalent to the Newton constant at current time\footnote{One may, in principle, consider to relax this condition and allow for slight departures from $G$ \cite{Martins17}. This would ensure that $F(R)$ exactly recovers the $\Lambda$CDM behaviour at large curvatures \cite{CMB limit}.} \cite{Dick04,Dominguez04}. This implies:
\begin{equation}
F\big|_{z=0}=1\ , \hspace{0.5cm} F'\big|_{z=0}=0\ .
\end{equation}
In \Cref{fig:F(R) Pade}, we show the behaviour of $F(R)$ obtained from the numerical solution of \Cref{eq:F-z} using the central values of \ref{cosm param Pade}.
To get $f(R)$, we then integrate this solution with the initial condition provided by evaluating \Cref{eq:Friedmann} at the present time:
\begin{equation}
f_0=6H_0^2(\Omega_{m0}-1)+R_0\  ,
\end{equation}
where $R_0$ is the present value of the Ricci scalar. We soon find that the analytical function matching the numerical solution is
\begin{equation}
f(R)_\text{Pad\'e}=a+bR^n\ ,
\label{eq:best f(R) Pade}
\end{equation}
where 
\begin{equation}
(a,\ b,\ n)=(-1.627,\ 0.866,\ 1.074)\ .
\end{equation}
We finally show in \Cref{fig:f(R) Pade} the Pad\'e reconstruction of the $f(R)$ function.
If we also take into account the upper and lower $1\sigma$ bounds of \ref{cosm param Pade}, we find the following intervals:
\begin{equation}
\left\{
\begin{aligned}
&a\in[-1.627,\ -1.326] ,\\
&b\in[0.733,\ 0.951] , \\
&n\in[1.025,\ 1.123] .
\end{aligned}
\right.
\label{bounds Pade}
\end{equation}

\begin{figure}
\begin{center}
\includegraphics[width=3.2in]{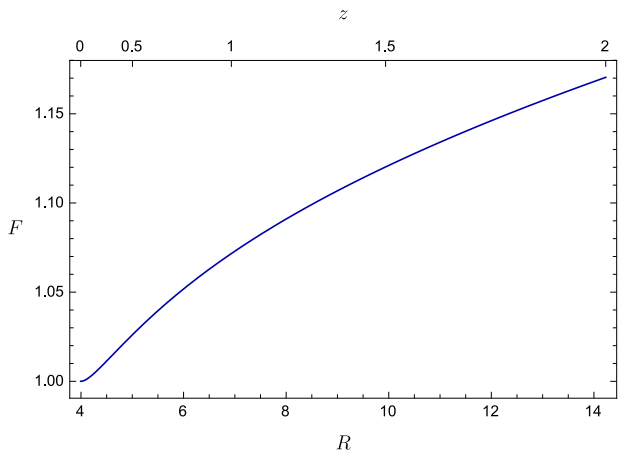}
\caption{Rational Chebyshev reconstruction of $F(R)$ in the redshift interval $[0,2]$.}
\label{fig:F(R) Chebyshev}
\end{center}
\end{figure}

\begin{figure}
\begin{center}
\includegraphics[width=3.2in]{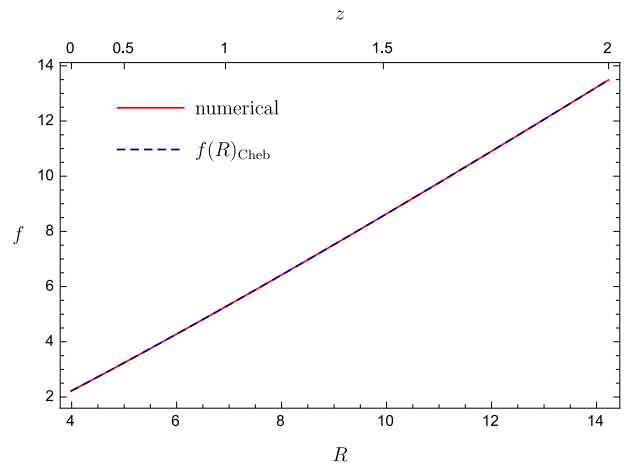}
\caption{Comparison between the rational Chebyshev reconstruction of $f(R)$ and its best analytical approximation (cf. \Cref{eq:best f(R) Cheb}).}
\label{fig:f(R) Chebyshev}
\end{center}
\end{figure}

On the other hand, for the rational Chebyshev approximation we use \cite{D'Agostino18}:
\begin{equation}
\left\{
\begin{aligned}
&h_0=0.6495^{+0.0189}_{-0.0194}\ , \\
&q_0=-0.278^{+0.021}_{-0.021}\ , \\
&j_0=1.585^{+0.497}_{-0.914}\ , \\
&s_0=1.041^{+1.183}_{-1.784}\ .
\end{aligned}
\right.
\label{cosm param Cheb}
\end{equation}
Adopting an analogous procedure as in the above case, we reconstruct $F(R)$ (see \Cref{fig:F(R) Chebyshev}) using the central values of \ref{cosm param Cheb}.
In this case, the numerical solution for $f(R)$, got from the integration of $F(R)$, matches with
\begin{equation}
f(R)_\text{Cheb}=\alpha +\beta R^m\ ,
\label{eq:best f(R) Cheb}
\end{equation}
where
\begin{equation}
(\alpha,\ \beta,\ m)=(-1.332, \ 0.749, \ 1.124)\ .
\end{equation}
Finally, the rational Chebyshev reconstruction of the $f(R)$ function is shown in \Cref{fig:f(R) Chebyshev}.
Considering the lower and upper $1\sigma$ bounds, we get the following intervals:
\begin{equation}
\left\{
\begin{aligned}
&\alpha\in[-1.481,\ -1.332] ,\\
&\beta\in[0.749,\ 0.818] , \\
&m\in[1.096,\ 1.124] .
\end{aligned}
\right.
\label{bounds Cheb}
\end{equation}

In both Pad\'e and Chebyshev cases, the reconstructed Palatini $f(R)$ action contains an explicit cosmological constant. Making use of  \ref{bounds Pade} and \ref{bounds Cheb}, it is possible to convert the bounds on the constants $a$ and $\alpha$ into bounds on the density parameter associated to $\Lambda$. We thus find:
\begin{align}
&\Omega_\Lambda^\text{Pad\'e}\in [0.711,0.748]\ , \\
& \Omega_\Lambda^\text{Cheb}\in [0.723,0.742]\ .
\end{align}
The above $1\sigma$ constraints do not include the specific value $\Omega_\Lambda=0.7$ expected since the assumption of a flat universe with $\Omega_{m}=0.3$. This is due to the degrees of approximation of the rational polynomials employed in the cosmological analysis. In fact, one should bear in mind that the reconstructed technique we propose may be further refined by increasing the order of polynomials to make the predictive power of the approximations more effective up to the desired level.

\subsection{Comparison with the concordance model}

\begin{figure}
\begin{center}
\includegraphics[width=3.2in]{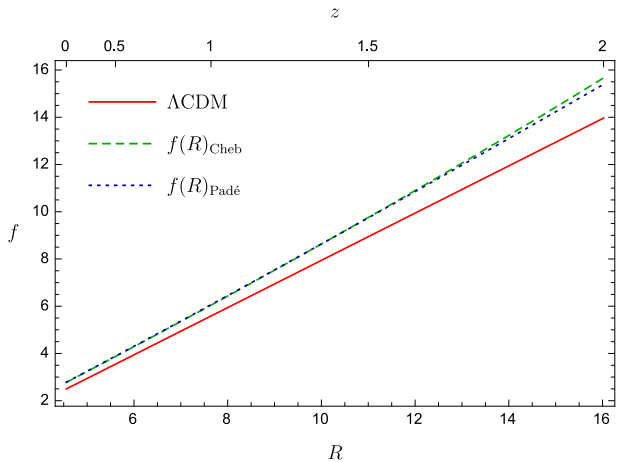}
\caption{Comparison of the $f(R)$ action of $\Lambda$CDM with the Pad\'e and the rational Chebyshev reconstructions.}
\label{fig:f(R) comparison}
\end{center}
\end{figure}

\begin{figure}
\begin{center}
\includegraphics[width=3.2in]{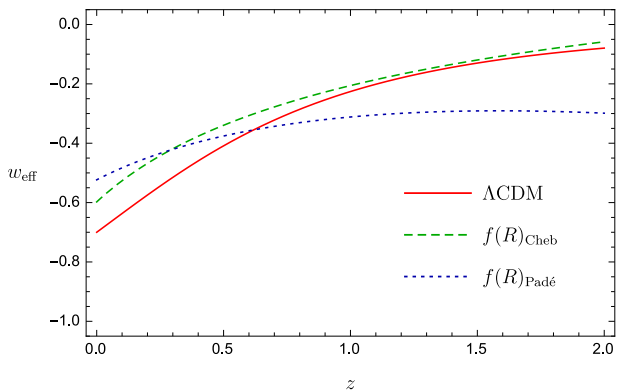}
\caption{Comparison of the effective equation of state parameter of $\Lambda$CDM with the Pad\'e and the rational Chebyshev reconstructions.}
\label{fig:w_eff comparison}
\end{center}
\end{figure}

It is interesting to compare our results with the cosmological predictions of the standard $\Lambda$CDM model.
We show in \Cref{fig:f(R) comparison} the comparison between the concordance $\Lambda$CDM action with $\{h_0, \Omega_{m0}\}=\{0.7, 0.3\}$ and the best reconstructions of $f(R)$ we obtained through rational approximations.
Moreover, from $H(z)$ one can calculate the effective equation of state parameter in terms of the redshift as
\begin{equation}
w_\text{eff}(z)=-1+\dfrac{2}{3}(1+z)\dfrac{H'(z)}{H(z)}\ .
\end{equation}
\Cref{fig:w_eff comparison} shows the comparison between the different models as result of using the best-fit values for the cosmographic parameters.
The functions $f(R)$ got from Pad\'e and Chebyshev polynomial reconstructions show shapes similar to the $\Lambda$CDM model. This fact seems to differ at the level of the equations of state. In fact, if one compares the Chebyshev and Pad\'e reconstructions of the equation of state with respect to the concordance paradigm, it seems that the three approaches quite differ from each other. This represents a consequence of the non-linearity of our $f(R)$, i.e. small variations in the Lagrangians may give significant variations over the equation of state.

For the sake of clarity, it is necessary to stress that our choice for the set of cosmographic parameters seems to rule out $\Lambda$CDM at the very beginning. We however take numerical outcomes which have been evaluated at $1\sigma$ levels only. Such results provide a significance which cannot exclude the concordance model \emph{a priori}. We thus conclude that our analyses may indicate our rational expansions suffer from propagation of non-linearities. In such a way, one can imagine a weak model-dependence of Pad\'e and Chebyshev polynomials. The problem is however solvable by considering refined analyses or by combining additional techniques of rational expansions. Indeed, with these guidelines we notice  rational polynomials in cosmographic frameworks become much more efficient tools to constrain background cosmology as higher orders are involved, i.e. as non-linearities are healed with refined analyses making use of combinations of rational expansions. We thus believe the initial choice over cosmographic coefficients does not significantly influence the whole cosmographic treatments and the concordance model is not \emph{a priori} ruled out.

\subsection{Comparisons with previous findings}

Previous approaches applied to cosmography have been commonly investigated in the framework of the metric formalism. Higher departures with our results may be found in the framework of the $f(R)$ reconstructions. In particular, the functional form of $f(R)$ has been showed to be much more complicated in the metric formalism with respect to Palatini \cite{primapires1,primapires2}. This may be due to the fact that here field equations are second order differential equations. In other words, passing from fourth to second order, one immediately gets that the whole equations should be much more similar to standard Einstein's gravity at the infrared regime. For these reasons, we did not require \emph{a priori} that our test-functions reduce directly to $f(R)\sim R$. On the other side, in \cite{Pires10} the authors proposed to adopt cosmography in order to quantitatively figure out which classes of viable models in the Palatini formalism are effectively able to describe the universe dynamics. They developed it by studying the evolution of $q$ in terms of the redshift. Further, they assumed the existence of a past matter-dominated epoch, ruling out the entire branch of negative values inside the parametrization $f\sim R-\beta R^{-n}$. Those results, however, are the direct consequence of postulating an $f(R)$ function that reduces to $\Lambda$CDM via the limit $f(R)\sim R+G(R)$, with $G(R)$ the cosmographic correction.
Their prescription is here confirmed by our bounds got for both Chebyshev and Pad\'e polynomials. Both the approaches, i.e. making use of simple cosmography with the recipe of a matter-dominated phase and ours, are intertwined and confirm that in the case of Palatini gravity the form of $f(R)$ is slightly departing from standard Einstein's gravity.

Earlier studies of Palatini $f(R)$ cosmology considered models such as $f(R) =\beta R^n$ or $f(R) =\alpha \ln R$ to explore departures from GR. These models were proved to be successful in fitting data of SN Ia, BAO and gas mass fraction in galaxy clusters \cite{Capozziello06,Borowiec06}, albeit no improvement with respect to the standard $\Lambda$CDM model emerged from the statistical analysis. Tight constraints on the class of models $f(R)=R-\alpha R^{\beta}$ were obtained by combining the SN Ia and BAO data with the CMB shift parameter and the evolution of linear perturbations \cite{Koivisto06,Amarzguioui06}. The allowed interval for $\beta$ was found to be $\sim 3\times 10^{-5}$ around $\beta=0$, with $\alpha$ similar to the cosmological constant. These constraints become even more stringent when taking into account the predictions of the CMB and matter power spectra, which confined the allowed parameter space to a tiny region around $\Lambda$CDM \cite{Li07}. Such results were later confirmed using updated data \cite{Carvalho08,Santos08} and strong lensing \cite{Ruggiero09,Yang09}. Thus, all the previous studies obtained by employing a specific parametrization suggest that the Palatini $f(R)$ theories are hardly distinguishable from $\Lambda$CDM. Similar conclusions can be drawn from our model-independent analysis.

Finally, Palatini $f(R)$ cosmology was also studied via a Noether symmetry approach in \cite{Roshan08}. Using the dynamical equivalence between $f(R)$ gravity and scalar-tensor theories, it was shown that the Noether symmetry always exists for $f(R)\sim R^n$ in the case of a matter-dominated universe. It is interesting to note that, in the matter-dominated universe, while in the metric formalism this symmetry exists only for $f(R)\sim R^{3/2}$ \cite{Capozziello08}, in the Palatini approach the Noether symmetry exists for Lagrangians with any arbitrary power of $R$.

\subsection{Dependence on $\Omega_{m0}$}

Current matter density cannot be obtained \emph{directly} from cosmographic analyses. In several cases it turns out to be highly model-dependent. In agreement with this, the constructed Pad\'e and Chebyshev expansions of the luminosity distance do not depend upon $\Omega_{m0}$, but rather on cosmographic parameters only. This is why we here assumed to fix matter density by $\Omega_{m0}=0.3$, which is consistent at the $1\sigma$ level with Planck's results, i.e.  $\Omega_{m0}=0.308 \pm 0.012$ \cite{Planck15}.
For the sake of clearness, this procedure is statistically disfavored with respect to let $\Omega_{m0}$ free to vary. To better motivate our choice and to check possible dependence of our outcomes over $\Omega_{m0}$, we take a spread interval for $\Omega_{m0}$: $\Omega_{m0}\in[0.25,0.35]$. We chose such an interval in agreement with the most recent bounds got from current experimental analyses. Within the considered range, our numerical procedure provides the same functional form of $f(R)$, as in Eqs. \Cref{eq:best f(R) Pade} and \Cref{eq:best f(R) Cheb}, with a maximum relative difference for the coefficients of about $\leq9\%$ at the bounds of the interval. This confirms both the goodness of our method and the stability of our results.

%%%%%%%%%%%%%%%%%%%%%%%%%%%%%%%%%%%%%%%%%%%%%%%%%%%%%%%%%%%

\section{Final remarks}
\label{sec:conclusion}

Among several possibilities, the promising paradigm of Palatini $f(R)$ gravity can be used to account for the universe's acceleration at late times. We considered the Palatini approach and we wondered how to frame $f(R)$ without postulating the model at the very beginning. Indeed, as in the standard metric approach, the $f(R)$ form is unknown \emph{a priori}. To find out possible clues toward its determination, one may postulate the form of both the pressure and density. Although appealing this strategy does not provide a model independent method to trace out $f(R)$ at late times. To heal this issue we took the energy momentum tensor free from any \emph{ad hoc} assumptions. We only considered a homogeneous and isotropic universe in which one can expand the luminosity distance around $z\simeq0$. We thus reconstructed the Palatini $f(R)$ in a model-independent fashion, by framing the cosmic expansion history through rational approximations of the luminosity distance. We chose rational series in order to characterize optimal convergence properties at higher-redshift domains, i.e. for $z\leq2$. In particular, we specifically considered the $(2,2)$ Pad\'e and the $(2,1)$ rational Chebyshev polynomials. We chose such orders since they have been proven to significantly reduce the error propagation on estimating the cosmographic series.

\noindent Adopting the most recent bounds on the cosmographic parameters, we built up accurate approximations of Hubble's rate up to $z\simeq 2$.
Using $H(z)$ and $\frac{dH}{dz}$, we immediately found $F\equiv df/dR$ in function of $z$ only. Thus, we numerically inverted $R(z)$ and plugged it back in $F(z)$ to finally reach the form of $f(R)$. We portrayed the evolutions of $f(z), R(z)$ and $f(R)$ at different redshift domains, and alternatively in terms of the Ricci curvature. From our outcomes, we found that best analytical matches to the numerical solutions are $f(R)_\text{Pad\'e}=a+bR^n$ where $(a, b, n)=(-1.627, 0.866, 1.074)$, and $f(R)_\text{Cheb}=\alpha+\beta R^m$ where $(\alpha, \beta, m)=(-1.332,  0.749,  1.124)$.
Our analyses showed small deviations from the concordance $\Lambda$CDM model based on GR. This has been better confirmed by checking the behaviour of the effective equation of state parameter, here evaluated for the above sets of parameters. We also underlined specific differences between our approach and previous ones applied to cosmography. Higher departures with our results may be found in the framework of the $f(R)$ reconstructions. In particular, the functional form of $f(R)$ has been got to be much more complicated in the metric formalism with respect to the Palatini one. We interpreted this by the fact that here Palatini's field equations are second order differential equations. In other words, passing from fourth to second order, one immediately gets that the whole equations should be much more similar to standard Einstein's gravity at the infrared regime. Our results also confirmed that the exponents in the term $\sim R^n$ should be positive.

\noindent Future developments will be devoted to apply our rational approximants to higher redshift domains. In such a way we will realize whether $f(R)$ will be much more complicated to characterize other epochs of the universe evolution. We also will study the consequences of our cosmographic Palatini $f(R)$ model with the Cosmic Microwave Background, showing how the power spectrum would be influenced by our predictions.

\section*{Acknowledgements}

This paper is based upon work from COST action CA15117 (CANTATA), supported by COST (European Cooperation in Science and Technology).

%\clearpage

\appendix*
\begin{widetext}
\section{Rational approximations of the luminosity distance}

We here report the $(2,2)$ Pad\'e and the $(2,1)$ rational Chebyshev approximations of $d_L(z)$, respectively:
\begin{align}
 P_{2,2}(z)&=\dfrac{1}{H_0}(6 z (10 + 9 z - 6 q_0^3 z + s_0 z - 2 q_0^2 (3 + 7 z) - q_0 (16 + 19 z) +
     j_0 (4 + (9 + 6 q_0) z))\Big/(60 + 24 z + 6 s_0 z - 2 z^2 \nonumber \\	
 &	 + 4 j_0^2 z^2 - 9 q_0^4 z^2 - 3 s_0 z^2 + 6 q_0^3 z (-9 + 4 z) + q_0^2 (-36 - 114 z + 19 z^2)\ ,
 \end{align}

 \begin{align}
 R_{2,1}(z)&=\dfrac{1}{H_0}(-((3 (16 (-1 - j_0 + q_0 + 3 q_0^2) (7 - j_0 + q_0 + 3 q_0^2) - (18 + 5 j_0 (1 + 2 q_0) - 3 q_0 (6 + 5 q_0 (1 + q_0)) + s_0) (14 \nonumber \\
  &+ 5 j_0 (1 + 2 q_0) - q_0 (14 + 15 q_0 (1 + q_0)) + s_0)))/(14 + 5 j_0 (1 + 2 q_0) - q_0 (14 + 15 q_0 (1 + q_0)) + s_0))+4 (47 - j_0 \nonumber \\
  &+ q_0 + 3 q_0^2 - (12 (-1 + q_0) (1 + j_0 - q_0 (1 + 3 q_0)))/(14 + 5 j_0 (1 + 2 q_0) - q_0 (14 + 15 q_0 (1 + q_0)) + s_0)) z \nonumber \\
  &-(4 (12 (-1- j_0 + q_0 + 3 q_0^2) (7 - j_0 + q_0 + 3 q_0^2) + 4 (1 + j_0 - q_0 (1 + 3 q_0))^2 - (14 + 5 j_0 (1 + 2 q_0) - q_0 (14  \nonumber \\
  &+ 15 q_0 (1 + q_0))+ s_0)^2) (-1 + 2 z^2))/(14 + 5 j_0 (1 + 2 q_0) - q_0 (14 + 15 q_0 (1 + q_0)) + s_0))\Big/(192 (1 + (4 (1 + j_0  \nonumber \\
  &- q_0 (1 + 3 q_0)) z)/(
     14+5 j_0 (1 + 2 q_0) - q_0 (14 + 15 q_0 (1 + q_0)) + s_0))) \ .
 \end{align}

\end{widetext}

\end{document}